\documentclass{appolb}
\usepackage{epsfig}

\newcommand{\grsim}{\mbox{\raisebox{-0.6ex}{$\stackrel{>}{\sim}$}}\:}

\begin{document}
\title{Hydrodynamic Approaches to\\
Relativistic Heavy Ion Collisions%
\thanks{Presented at XXXIV International Symposium
on Multiparticle Dynamics,
July 26 - August 1, 2004,
Sonoma State University, Sonoma County, California, USA }%
}
\author{Tetsufumi Hirano
\thanks{Present address: 
Department of Physics, Columbia University,
Pupin 925, 538 West 120th Street,
New York, NY 10027, USA
}
\address{RIKEN BNL Research Center, Brookhaven National
Laboratory,\\
Upton, NY 11973, USA}
}
\maketitle
\begin{abstract}
We give a short review
about the hydrodynamic model and its application
to the elliptic flow phenomena and the pion
interferometry
in relativistic heavy ion collisions.
\end{abstract}

\section{Introduction}

First data reported by the STAR Collaboration at 
RHIC \cite{STAR:v2} has a significant meaning that
the observed large magnitude of elliptic flow
for charged hadrons
is consistent with hydrodynamic predictions
\cite{KHHH}.
This suggests that
large pressure possibly in the partonic phase
is built at the early stage
($\tau \sim 0.6$ fm/$c$)
in Au+Au collisions at $\sqrt{s_{NN}} =$ 130 and 200 GeV.
This situation at RHIC is in contrast to
that at lower energies such as AGS or SPS
where hydrodynamics always overpredicts
the data \cite{NA49:v12}.
Moreover, this also suggests that the effect of the viscosity
in the QGP phase is remarkably small and that the QGP is 
almost a perfect fluid \cite{S}.
Hadronic transport models
are very good to describe experimental data
at lower energies, while they fail to reproduce
such large values of
elliptic flow parameter at RHIC (see, e.g., Ref.~\cite{BS}).
So the importance of hydrodynamics is rising
in heavy ion physics.
After the first STAR data were published \cite{STAR:v2}, 
other groups at RHIC have also obtained the data
concerning with flow phenomena \cite{Retiere}.

Contrary to the success of the hydrodynamics to describe
the elliptic flow,
many dynamical models including hydrodynamics cannot
reproduce the HBT radii \cite{STAR:HBT,PHENIX:HBT,magestro}.
It is known as the ``HBT puzzle".

To understand
these experimental data,
hydrodynamic analyses are performed
extensively \cite{QGP3:hydro:Pasi,QGP3:hydro:Peter}.
In this short review, we highlight several
results mainly on elliptic flow and on the HBT radii
from hydrodynamic calculations.

\section{Basics of Ideal Hydrodynamics}


Hydrodynamic equations represent the
energy-momentum conservations
$\partial_\mu T^{\mu\nu}=0$.
In the ideal hydrodynamics, the energy-momentum tensor 
becomes $T^{\mu\nu} = (e+P) u^\mu u^\nu - Pg^{\mu\nu}$,
where $e$ is the energy density, $P$ is the pressure,
and $u^\mu$ is the four fluid velocity.
When there are conserved quantities 
such as the baryon number or the number of chemically frozen
hadrons,
one needs to solve
the continuity equations
$\partial_\mu n_i^\mu = 0$ together with the hydrodynamic equations.
In order to close the system of partial differential
equations, the equation of state (EoS)
$P(e,n_i)$ is needed.
The naive applicability conditions of ideal
hydrodynamics are that the mean free path among the
particles is much smaller than the typical size
of the system and that the system keeps local thermal equilibrium
during expansion.
From these conditions, one cannot use hydrodynamics
for initial collisions, final free streaming
and high $p_T$ ($\grsim$ 2 GeV/$c$) particles.
So one needs an interface between
the pre-thermalisation stage and the hydrodynamic
stage at the initial time.
Moreover,
the system eventually breaks up and 
cannot keep thermalisation
at the later stage.
This means a prescription of freezeout
is needed in the hydrodynamic model in evaluating
the particle spectra.
Therefore one needs another interface
between the hydrodynamic stage and
the free streaming stage.
 In what follows, we discuss particularly
equations of state, initial conditions
 and freezeout prescriptions used in the literature.

\subsection{Equation of State}

The main ingredient of the hydrodynamic model is
the equation of state (EoS) for thermalised
matter produced in heavy ion collisions.
Ideally, one uses the EoS taken from
the first principle
calculations of QCD, namely,
lattice QCD simulations \cite{Karsch}.
More practically,
one can use the resonance gas model for the hadron phase and
the massless free parton gas for the QGP phase.
By matching these two models at the critical temperature,
one obtains the first order phase transition model
with a latent heat $\sim$1 GeV/fm$^3$.
In the mixed phase at $n_B=0$, the sound velocity
$c_s^2 = {\partial P}/{\partial e}$
is vanishing.
Recent lattice QCD simulations
tell us that the phase transition
seems to be crossover in vanishing baryonic chemical
potential.
Discontinuity of the thermodynamic
variables does not exist in the crossover
phase transition.
It should be emphasized, however, that
the energy density and the entropy density increase
more rapidly than the pressure
in the vicinity of
the phase transition region $\Delta T \sim 0.1 T_c$.
This also leads to very small sound velocity
near the phase transition region.
Therefore it is very hard in general to find flow observables which
distinguish the crossover phase transition with
a rapid change of the thermodynamic variables
from the first order phase transition.


\subsection{Initial Condition}

Once initial conditions are assigned, one can numerically simulate
the space-time evolution
of thermalised matter 
which is governed by hydrodynamic equations.
Usually, initial conditions are parametrised
based on some physical assumptions.
Transverse profile of the energy/entropy
density is assumed to be proportional to
the number density of participants $\rho_{\mathrm{part}}$,
the number density of binary collisions $\rho_{\mathrm{coll}}$
or linear combination of them.
Initial transverse flow is usually taken to be vanishing.

On the other hand, one can introduce
model calculations
to obtain the initial condition
of hydrodynamic simulations.
Event generators can be used
to obtain the energy density distribution
at the initial time.
Recently, the SPheRIO group
employs an event generator NeXus
and takes an initial condition
from this model
in the event-by-event basis \cite{OAHK,HAMA}.
The resultant energy density
distribution in the transverse
plane has highly bumpy structures \cite{OAHK,GRZ}.
Smooth initial conditions 
used in the
conventional hydrodynamic simulations
are no longer expected in one event.
Another important example
which is relevant
at very high collision energies
is an initial condition
taken from the Colour Glass Condensate
picture. See Ref.~\cite{HiranoNara6} for recent calculations.

\subsection{Freezeout}
\label{sec:freezeout}

Conventional prescription to obtain the invariant momentum
spectra
from the hydrodynamic simulations
is to employ the Cooper-Frye formula
\cite{CF}.
%
The physical picture described by the Cooper-Frye
formula
is sometimes called ``sudden freezeout" since
the mean free path
is suddenly changed from zero to infinity
through a thin freezeout hypersurface.
Instead of using this,
one can use a hadronic cascade model
to describe the space-time
evolution of hadrons \cite{BD,TLS}.
The mean free path among hadrons
is finite and depends on hadronic species.
Hence, 
one can describe a continuous
freezeout picture
through hadronic transport models.
Note that continuous
particle emission can be considered
within the hydrodynamics \cite{Grassi:1994nf}.
Adoption of hadronic transport models
after hydrodynamic evolution of the QGP liquid
could refine the dynamical modeling
of relativistic heavy ion collisions.
However, it is not so easy
to connect them in a systematic and proper way \cite{Bugaev}.

\section{Hydrodynamic Results for $v_2$}
\label{sec:v2}

\begin{figure}[t]
\begin{center}
\includegraphics[width=0.4\textwidth]{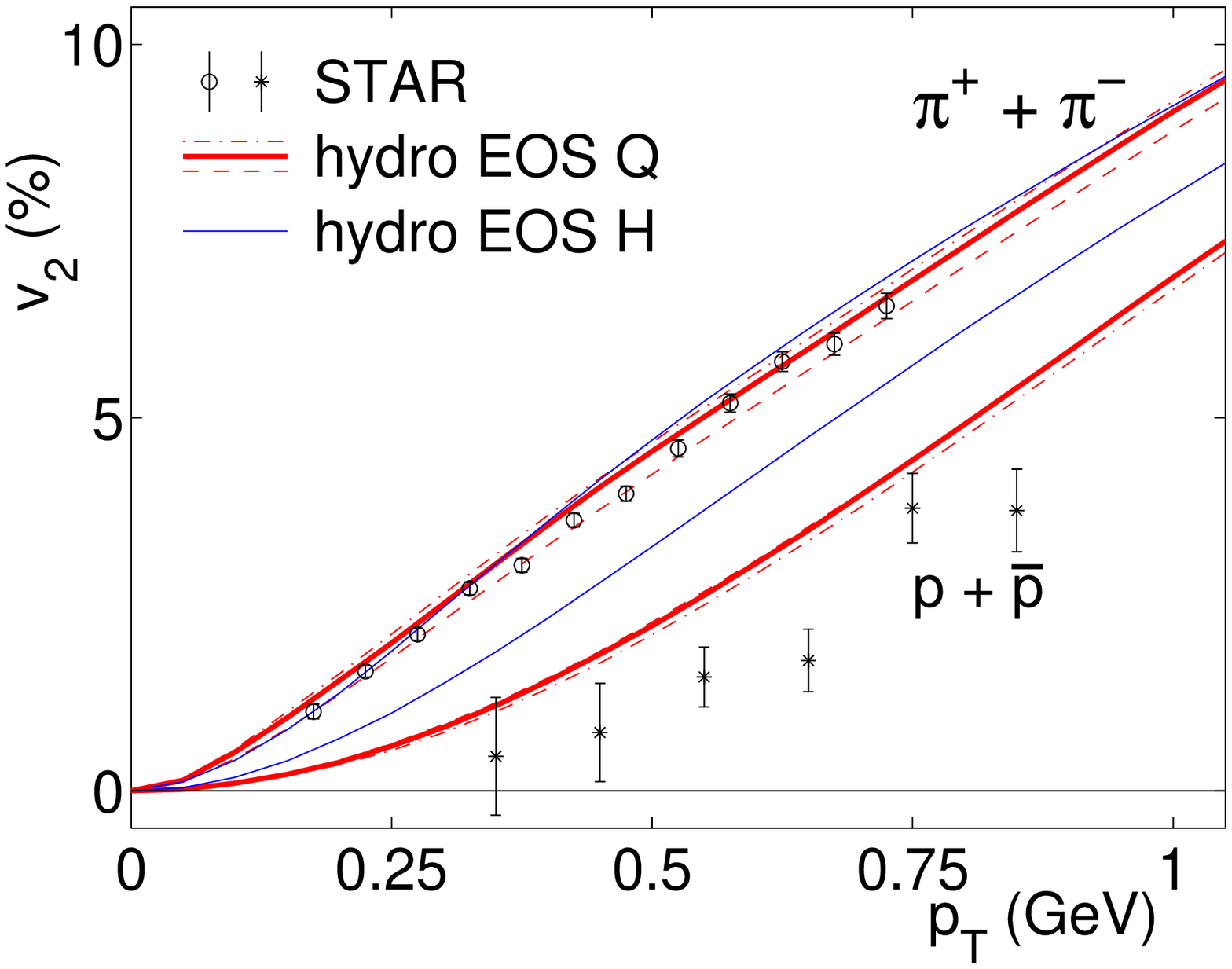}
\includegraphics[width=0.5\textwidth]{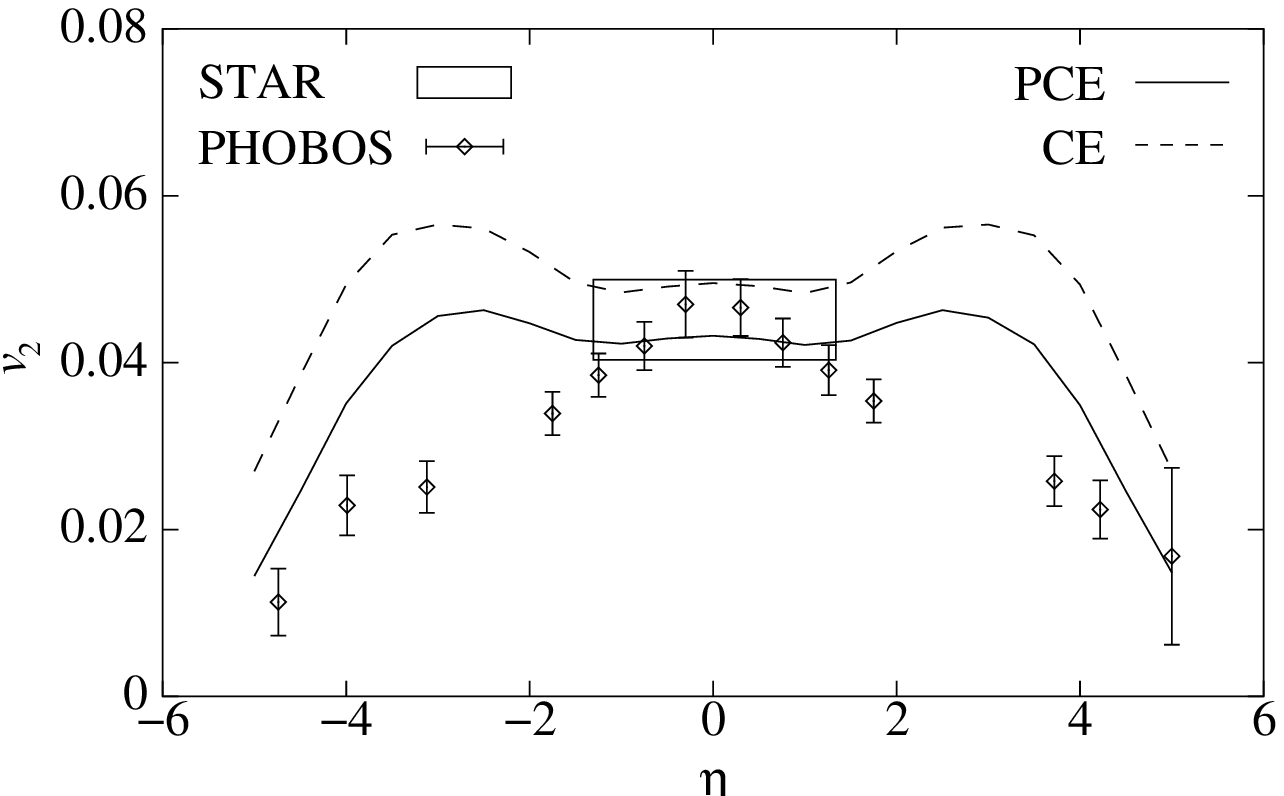}
\end{center}
\caption{
(Left)
Transverse momentum dependence of $v_2$ for pions and (anti)protons
\cite{QGP3:hydro:Peter} are compared with the STAR data \cite{STAR:v2id}
in minimum bias Au+Au collisions at $\sqrt{s_{NN}}=130$ GeV.
EoS Q stands for the first order
phase transition model, while EoS H stands
for the resonance gas model.
(Right)
$v_2$ for charged hadrons
as a function of pseudorapidity $\eta$
in Au+Au collisions at $\sqrt{s_{NN}}=130$ GeV \cite{STAR:v2,PHOBOS:v2eta}.
PCE means the EoS of partial chemical equilibrium,
whereas CE means the EoS of chemical equilibrium.
Figure is taken from Ref.~\cite{HiranoTsuda}.
}
\label{fig:v2pt}
\end{figure}

Assuming the Bjorken flow \cite{BJ}
for the longitudinal direction,
one can solve the hydrodynamic equations
only in the transverse plane at midrapidity.
Systematic studies based on this (2+1)-dimensional hydrodynamic
model are performed in Ref.~\cite{KHHH}.
For the EoS, complete chemical equilibrium is assumed for both
the QGP phase and the hadron phase.
$p_T$ dependences of $v_2$ for pions and protons 
from this model \cite{QGP3:hydro:Peter} are
compared with the STAR data \cite{STAR:v2id}
in Fig.~\ref{fig:v2pt} (left).
By employing the EoS with phase transition,
the hydrodynamic model correctly
reproduces $v_2(p_T)$ and its mass-splitting behavior
below $p_T=1$ GeV/$c$.
On the other hand, $v_2(p_T)$ for (anti)protons
from the resonance gas model
does not agree with the data.
Although the reason for the difference
of the result between these two EoS models
is not so clear,
the experimental data favors
the QGP EoS.
Due to the assumption of chemical equilibrium
in the hadron phase, 
this model does not reproduce particle ratio
and spectra simultaneously.
 It is of importance to study whether the agreement
 with the experimental data still holds
 even when the assumption of chemical equilibrium
 in the hadron phase
 is abandoned \cite{HiranoTsuda}.


One needs a full 3D
hydrodynamic simulation 
to obtain the rapidity dependence of $v_2$.
First analysis of $v_2(\eta)$ at RHIC based on
the full 3D hydrodynamic model
is performed in Ref.~\cite{HiranoTsuda,Hirano}. 
Figure \ref{fig:v2pt} (right) shows $v_2(\eta)$ for charged hadrons
in Au+Au collisions at $\sqrt{s_{NN}}=130$ GeV \cite{STAR:v2,PHOBOS:v2eta}.
Here the initial condition for the energy density is so
chosen as to reproduce the pseudorapidity distribution
of charged hadrons.
Hydrodynamic results are consistent with
the experimental data only near the midrapidity
while hydrodynamics overpredicts
the data in the forward/backward rapidity regions.
Multiplicity is not so large in the forward/backward
rapidity regions, so equilibration of the system
tends to be spoiled.

\begin{figure}[t]
\begin{center}
\includegraphics[width=0.25\textwidth]{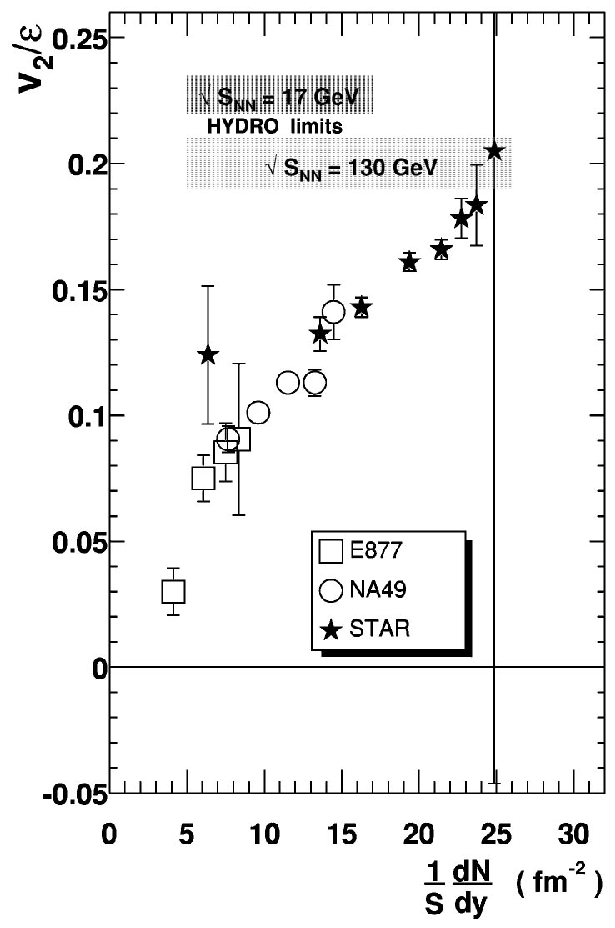}
\includegraphics[width=0.45\textwidth]{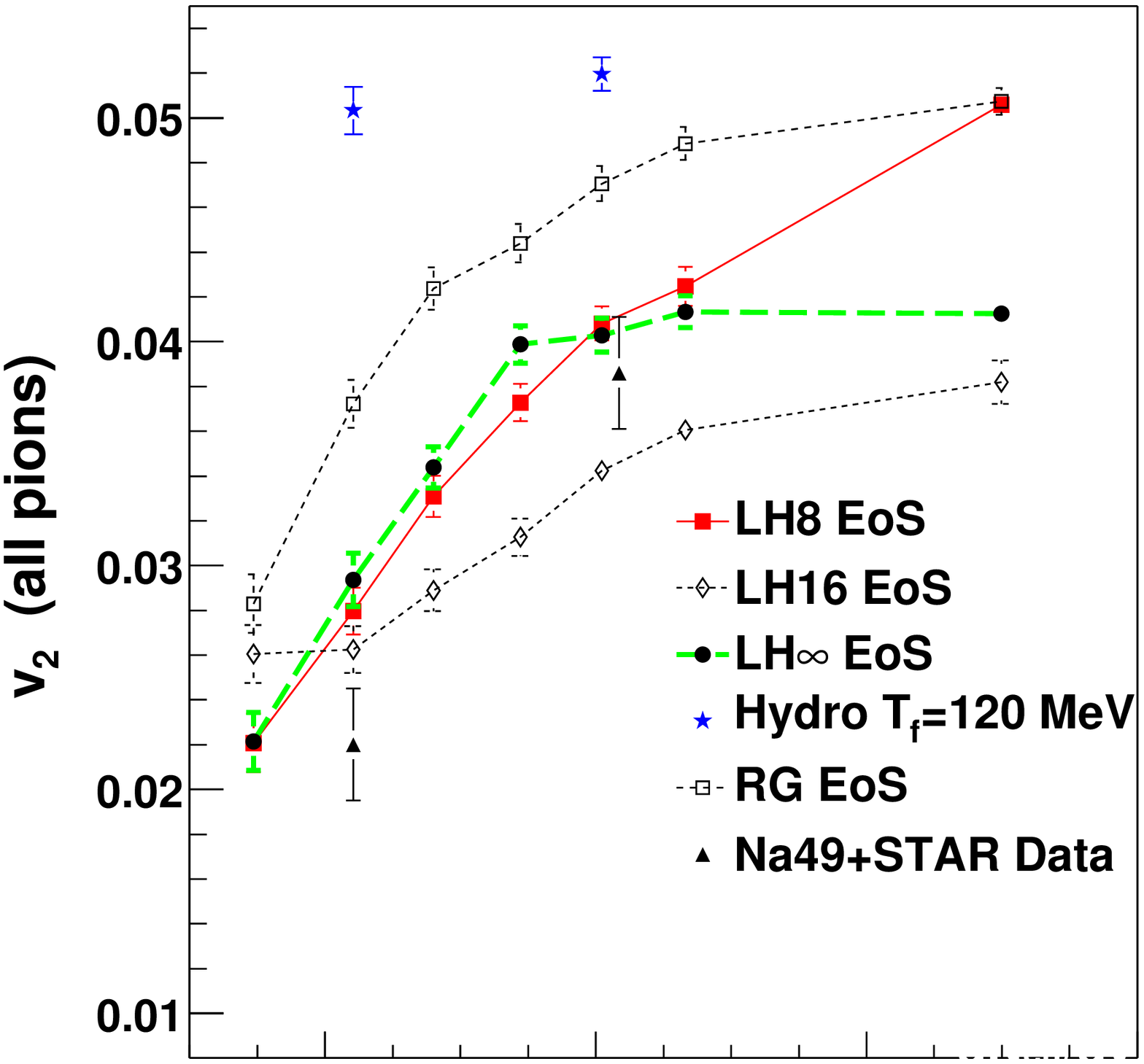}
\end{center}
\caption{
(Left) Excitation function of $v_2/\varepsilon$.
Figure taken from Ref.~\cite{STAR:scaledv2}.
(Right) Excitation function from the hydro+cascade model.
Figure taken from Ref.~\cite{TLS}.
}
\label{fig:v2overeps}
\end{figure}

Figure \ref{fig:v2overeps} (left) shows the excitation
function of $v_2$ compiled
by the STAR Collaboration \cite{STAR:scaledv2}.
Hydrodynamic results presented in this
figure are based on the same model
discussed in Fig.~\ref{fig:v2pt} (left).
Data points continuously
increase with the unit rapidity density
at the AGS, SPS and RHIC energies.
However, the hydrodynamic response $v_2/\varepsilon$
is almost flat or slightly decreases
with the multiplicity.
The data points seem to reach the ``hydrodynamic limit"
for the first time at the RHIC energy.
The deviation between the hydrodynamic results and
the data plots below $(1/S)dN/dy = 25$
reminds us the pseudorapidity dependence of elliptic flow
in Fig.~\ref{fig:v2pt} (right). 
The deviation might come from a common origin \cite{Heinzqm2004}:
The small multiplicity both in forward rapidity region
at the RHIC energy
and at midrapidity at the lower collision energies
could cause the partial thermalisation.
In the low density limit, $v_2$ is actually
proportional to the number density \cite{Heiselberg}.
The shape of $v_2(\eta)$ data in forward rapidity region
looks similar to that of
the pseudorapidity distributions \cite{Steinberg}.
Similarly, data plots of excitation function increase
almost linearly with the particle density.
These results suggest that thermalisation
is not achieved completely in forward rapidity region
at the RHIC energy and at midrapidity at the SPS energies.


Figure \ref{fig:v2overeps} (right) shows
the excitation function from the hydro+cascade
(RQMD) model \cite{TLS}
with Bjorken longitudinal flow \cite{BJ}.
Contrary to the conventional hydrodynamic models,
freezeout processes are automatically
described by the cascade model without any adjustable parameters.
The excitation function in the case of 
the latent heat $\sim 0.8$ GeV/fm$^3$
linearly increases
with the multiplicity, which is consistent with
the experimental data.
The difference from the conventional hydrodynamic results
might come from the strong viscosity in the hadron phase
within the cascade calculation.
It should be noted that $v_2(p_T)$, its mass dependences
and particle spectra/ratio at midrapidity
are also reproduced by this hybrid model \cite{TLS}.

\section{Hydrodynamic Results for HBT Radii}

Although many hydrodynamic calculations are already performed,
one does not succeed to interpret the HBT puzzle yet.
The main reason why the hydrodynamic simulations
overestimate the $R_{\mathrm{out}}/R_{\mathrm{side}}$
is the negative value of
the correlation between $\tilde{x}_{\mathrm{out}}$ and $\tilde{t}$
which comes from the hydrodynamic source
function~\cite{QGP3:hydro:Peter}.
Here $\tilde{x}=x-\langle x \rangle$ and 
the average is taken over the source function.
Note that positive $\tilde{x}_{\mathrm{out}}$-$\tilde{t}$ correlation
can be obtained dynamically in a transport model \cite{LinKoPal}.
So detailed comparison of this result with hydrodynamic results would
be important in understanding the space-time evolution
of matter in heavy ion collisions.

As discussed in Sec.~\ref{sec:freezeout}, the conventional
hydrodynamics assumes the sudden freezeout which must be far from
the realistic situation. The HBT radii reflect the distribution of
the final scattering points. So realistic treatment of the
final decoupling stage is mandatory.~\footnote{The
effect of final multiple scattering
on the HBT radii is recently discussed in Refs.~\cite{CYWong,Kapusta}.
According to these analyses, the distribution
one can obtain from the HBT analysis might be the initial effective
one rather than the one of final scattering points.}
To remove this unwanted feature in the hydrodynamic model,
the continuous particle emission is proposed
in Ref.~\cite{HAMA,Grassi:1994nf}.
This prescription naturally gives that larger momentum particles
comes from the earlier stage and that smaller momentum particles
comes from the later stage.
For details of the results, see Ref.~\cite{HAMA}.

\begin{figure}[t]
 \begin{center}
 \includegraphics[width=0.4\textwidth]{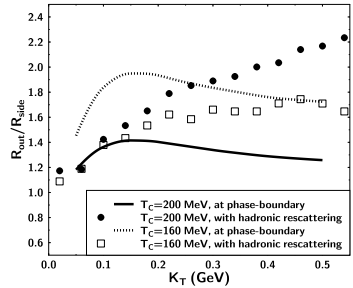}
 \end{center}
 \caption{$R_{\mathrm{out}}$/$R_{\mathrm{side}}$
 at hadronization (lines) and at freezeout (symbols)
 from a hydro+cascade approach.
 }
 \label{fig:HBT}
 \end{figure}

As shown in the previous section, the hydrodynamics
with a hadronic
cascade model at the late stage
reproduces the momentum space of particle
spectra from SPS to RHIC energies.
This indicates the dilute hadronic
gas should not be described by the ideal hydrodynamics.
In Ref.~\cite{Soff:2000eh}, a hydro+cascade (UrQMD) approach
is employed to calculate the ratio of the pion HBT radii
as shown in Fig.~\ref{fig:HBT}.
In the case of $T_c=160$ MeV,
the ratio becomes 2.0 around pair transverse
momentum $K_{T}=$0.15 GeV/$c$ at hadronization which is much larger than
the experimental data $\sim 1$ \cite{STAR:HBT,PHENIX:HBT}.
Although smearing of freezeout hypersurface
by using a hadronic cascade reduces the ratio to 1.4-1.6,
it is not enough to reproduce the experimental data.

\section{Summary and Outlook}

The most successful hydrodynamic approach
to elliptic flow
in relativistic heavy ion collisions is the hybrid model in which
the QGP phase is described by
the ideal hydrodynamics
while the hadron phase is described by a
hadronic cascade.
However, even within this model, the ratio of the HBT radii
is still larger than the data.
The Bjorken scaling solution is assumed 
in the current hydro+cascade models.
This means that current hybrid models
are available only near midrapidity.
Therefore it is desired to develop a model
in which a full 3D hydrodynamic
simulations combined with
a hadronic cascade model.
From agreement of excitation function
between the hybrid model and the experimental
data at midrapidity, the 3D hybrid model is expected to reproduce
the pseudorapidity dependence of elliptic
flow.
It should be emphasised again
that the hybrid model has its own problem
on the violation of energy momentum conservations
between the QGP liquid
and the hadron gas.
A hybrid simulation which incorporates a proper
treatment at the boundary between the QGP phase and
the hadron phase is now an open problem.

\end{document}